\documentclass[aip,graphicx]{revtex4-1}

\usepackage{graphicx}
\usepackage{dcolumn}
\usepackage{bm}

\usepackage[utf8]{inputenc}
\usepackage[T1]{fontenc}
\usepackage{mathptmx}
\usepackage{physics}
\usepackage{lipsum}
\usepackage{textcomp}

\draft 

\begin{document}


\title{Dissipation in Lagrangian formalism} 



\author{Andr\'as Szegleti}
\email[Corresponding author: ]{szegleti.a@gmail.com}
\affiliation{Department of Physics,
Budapest University of Technology and Economics,
H-1111 Budafoki \'ut 8., Budapest, Hungary}

\author{Ferenc M\'arkus}
\email[]{markus@phy.bme.hu}
\affiliation{Department of Physics,
Budapest University of Technology and Economics,
H-1111 Budafoki \'ut 8., Budapest, Hungary}


\date{\today}

\begin{abstract}
In this paper we present a method with which it is possible to describe a dissipative system in Lagrangian formalism, without the trouble of finding the proper way to model the environment. The concept of the presented method is to create a function that generates the measurable physical quantity, similarly to electrodynamics, where the scalar potential and vector potential generate the electric and magnetic fields.
\end{abstract}

\pacs{}

\maketitle 

\section{Introduction}
Newtonian mechanics can provide a general description of a physical system, as there are no limitations on the force terms, that can contribute to the equations of motion. This freedom is actually a drawback of the formalism, as correct equations of motion can be generated using ad-hoc forces, which limits the possibility of gaining predictions from the theory. 

Lagrangian mechanics is built on a more general principle, Hamilton's principle (or least action principle), which states, that there is a function \(L(q_i, \dot q_i, t)\), that describes the physical system, and the action functional
\begin{equation}
S = \int\limits_{t_1}^{t_2}\dd t L(q_i, \dot q_i, t)
\end{equation}
is extremal in case of physical trajectories. The equations of motion (Euler--Lagrange equations) for the system can be calculated from variational principle (functional derivative). This approach strongly limits the form of the equations one can derive using this formalism, which means that the form of Lagrangian describing a system is also restricted.

Dissipation, being a statistical phenomenon, could only be described by a Lagrangian containing all degrees of freedom (both for the system and its environment). Using this system-plus-reservoir approach, it is always an important question how the environment should be modelled. For example, using a harmonic bath model \cite{leggett1984quantum}, one assumes, the reservoir can be represented as a set of uncoupled harmonic oscillators. By using a different model for the environment, a set of two-state systems \cite{PhysRevB.48.13974,Prokof_ev_2000}, the resulting dynamics of the system might be different.

Using an explicitly time-dependent Lagrangian, the resulting equations of motion will show a dissipation of energy. This may give the idea that there might be a workaround with which it is possible to describe dissipation phenomena (more generally than dissipation of energy), without the trouble of finding the correct model for the environment. It has been the motive of several researches, considering both classical case and quantization throughout the years \cite{PhysRev.38.815, doi:10.1063/1.5022321, PhysRevA.61.022107}. One way of doing this (at least in principle), is to define a potential (by doubling the degrees of freedom) which generates the physical quantity.

\section{Introducing an abstract potential}
In this method, potential is a function which not necesarrily carries any physical meaning, but contains all physical information, thus it generates the observable physical quantity. It can be a purely mathematical tool to build Lagrangian formalism, that will generate the desired equation of motion for the observable in the end. It is possible to define a potential for any quantity described by any linear differential equation \cite{markus1991variational, gambar1991variational, nyiri1991construction}. The idea is analogous to how one deals with observables and potentials in electrodynamics. The observables (electric field and magnetic flux density) cannot be handled in Lagrangian formalism, but potentials (scalar potential and vector potential) can be defined, with corresponding differential equations of higher order.

\subsection{Creating a Lagrangian using potentials}
\label{sec:abstract-potential}
A general linear Euler--Lagrange equation can be written in the form
\begin{eqnarray}
\mathcal{\tilde D}\qty{\pdv{L}{\mathcal D\qty{u}}} = 0
\label{eq:EL-generalform}
\end{eqnarray}
where  $\mathcal{D}$ is a formal linear differential operator and $\mathcal{\tilde D}$ is its formal adjoint defined by
\begin{eqnarray}
\int\limits_\Omega\dd\tau\, v\cdot \mathcal{D} u - \int\limits_\Omega\dd\tau\, u\cdot \mathcal{\tilde{D}}v = \int\limits_{\partial \Omega}\dd\nu\,\mathcal B\qty{u, v},
\label{eq:lagrange-identity}
\end{eqnarray}
where $\mathcal B\qty{u, v}$ is called the bilinear concomitant. This definition provides the possibility to calculate $\mathcal{\tilde D}u$ through repeated integration by parts. Let's look at the example of a general differential operator of order $n$ acting on a function with a single variable
\begin{eqnarray}
\mathcal{D}u = p_n\dv[n]{u}{t}+p_{n-1}\dv[n-1]{u}{t} + \dots + p_1 \dv{u}{t} + p_0 u,
\end{eqnarray}
for which the adjoint operator acting on the function is
\begin{eqnarray}
\mathcal{\tilde D}u = (-1)^{n}\dv[n]{t}\qty(p_nu)+(-1)^{n-1}\dv[n-1]{t}\qty(p_{n-1}u) + \dots.
\end{eqnarray}
We can say that $\mathcal D$ is self adjoint if $\mathcal D u \equiv \mathcal{\tilde D}u$.

Suppose that a measurable physical quantity \(u(t)\) is described by the following inhomogeneous equation
\begin{eqnarray}
\mathcal D u(t) = c(t),
\label{eq:diffeq-orig}
\end{eqnarray}
where \(c(t)\) is arbitrary function. If \(\mathcal{D}\) is not self adjoint, it cannot be calculated from variational principle. 
One can define the potential $\phi(t)$ through the definition equation 
\begin{eqnarray}
u(t) = \mathcal{\tilde{D}}\phi(t).
\label{eq:def-eq-for-potential}
\end{eqnarray}
Substitute Eq.~(\ref{eq:def-eq-for-potential}) in Eq.~(\ref{eq:diffeq-orig}) and 
\begin{eqnarray}
\mathcal D \mathcal{\tilde D}\phi(t) = c(t).
\label{eq:EL-generalform-quadratic}
\end{eqnarray}
is received. By using Eq.~(\ref{eq:lagrange-identity}), it is easy to see that the differential operator \(\mathcal{D}' := \mathcal{D}\mathcal{\tilde{D}}\) is self-adjoint, hence the equation of motion for the potential \(\phi\) can be calculated from variational principle, so a Lagrangian exists, from which the equation of motion~(\ref{eq:EL-generalform-quadratic}) can be calculated. This Lagrangian can be written in the following form
\begin{eqnarray}
L = \frac{1}{2}\qty(\mathcal{\tilde D}\phi(t))\cdot\qty( \mathcal{\tilde D}\phi(t)) - \phi(t)\cdot c(t).
\end{eqnarray}
By using Eq.~(\ref{eq:EL-generalform}), the Euler--Lagrange equation can be calculated, resulting in Eq.~(\ref{eq:EL-generalform-quadratic})

\subsection{On the solutions}
The potential $\phi(t)$ contains all physical information, and some excess, non-physical information can be encoded in it as well. Consider the linear operator $\mathcal D$ and its adjoint $\tilde{\mathcal D}$, and suppose that $\mathcal D, \tilde{\mathcal D} \in \mathrm{Lin}\qty(V)$. As it is possible to obtain the original differential equation~(\ref{eq:diffeq-orig}) from Eq.~(\ref{eq:EL-generalform-quadratic}), it is safe to say that the kernel of $\tilde{\mathcal D}$ contains only non-physical information. By writing the solution $\phi(t)$ in such a way that $\phi(t) = \varphi(t)+\lambda(t)$, where $\tilde{\mathcal D}\varphi(t)\in \mathrm{Im}(\tilde{\mathcal D})$ and $\lambda(t)\in\mathrm{Ker}(\tilde{\mathcal D})$ it is easy to see that the $\lambda(t)$ term can be omitted
\begin{eqnarray}
\mathcal{D}u(t) =& \mathcal{D}\mathcal{\tilde D}\phi(t) = \mathcal{D}\mathcal{\tilde D}\qty(\varphi(t)+\lambda(t))  \nonumber \\
=& \mathcal D (\mathcal{\tilde D}\varphi(t)+  \mathcal{\tilde D}\lambda(t))= \mathcal D\mathcal{\tilde D}\varphi(t) .
\end{eqnarray}

This can be interpreted as a kind of gauge freedom, because by omitting the $\lambda(t)\in \mathrm{Ker}(\mathcal{\tilde D})$ part of the potential, the measurable physical quantity will stay invariant, so one can define the gauge transformation as
\[
\phi(x)\rightarrow \phi(x) + \Lambda(x)\qq{where }\Lambda(x)\in \mathrm{Ker}(\mathcal{\tilde D}).
\]

The solution of the adjoint equation $\tilde{\mathcal D}\lambda(t)=0$ is related to the time reversed process. Consider a homogeneous ordinary differential equation with constant coefficients, for which the differential operator is
\[
\mathcal{D} = \sum\limits_{n=0}^N p_n\dv[n]{}{t},
\]
The adjoint equation reads
\[
\mathcal{\tilde D}\zeta(t) = \sum\limits_{n=0}^N (-1)^n p_n\dv[n]{\zeta(t)}{t} = 0.
\]
It can be seen, that every odd order derivative changes its sign, and the even order terms are invariant. By changing the sign of the variable \(t\) (\(t\rightarrow -t\)), the adjoint equation can be rewritten
\[
\sum\limits_{n=0}^N (-1)^n p_n\dv[n]{\zeta(-t)}{t}\frac{1}{(-1)^n} = \sum\limits_{n=0}^N p_n\dv[n]{\zeta(-t)}{t} = \mathcal{D}\zeta(-t)=0.
\]
In such a simple case, it can be clearly seen, that if \(\zeta(t)\) is a solution of \(\mathcal{\tilde D}\zeta(t) = 0\), then its time reversed is a solution of \(\mathcal{ D}\zeta(-t) = 0\). As a consequence, \(\lambda(t)\in \mathrm{Ker} \mathcal{\tilde D}\) is related to the time reversed of \(v(t) \in \mathrm{Ker} \mathcal{D}\). Dissipative processes in nature tend to an equilibrium state, so the time reversed of these solutions are divergent. To obtain a stable solution, the divergent term (\(\lambda(t)\in \mathrm{Ker} \mathcal{\tilde D}\)) should be omitted.

\subsection{On initial conditions}
In theory it is easy to omit the solutions from $\mathrm{Ker}\mathcal{\tilde D}$ and for an analytical solution one can easily perform the correct gauge transformation. Unfortunately, it does not seem possible if we wish to solve the differential equation numerically. A good idea would be to choose initial and boundary value conditions carefully so that the non-physical part \(\lambda(t)\) vanishes. The aim is to find the relation between the initial conditions for the potential and the initial conditions for the measurable.

For the sake of simplicity, let's deal with only one variable. Firstly write the general solution for the inhomogeneous equation Eq.~(\ref{eq:EL-generalform-quadratic}) in the form
\begin{eqnarray}
\phi(t) = \sum\limits_{k=1}^N \qty[a_k\varphi_k(t) + b_k\lambda_k(t)] + \xi(t),
\label{eq:general-solution-for-potential}
\end{eqnarray}
where $\mathcal{\tilde D}\varphi_k(t)$ form the basis for the subspace $\mathrm{Ker}(\mathcal{D})$ and $\lambda_k(t)$ form the basis for the subspace $\mathrm{Ker}(\mathcal{\tilde D})$ and $\xi(t)$ is a particular solution of the inhomogeneous equation (so the solution \(\phi(t) = \varphi(t) + \lambda(t)\) is expanded on a basis). To solve a differential equation of order $2N$, we need $2N$ initial conditions. As the number of initial conditions and the number of coefficients (\(a_k\) and \(b_k\)) are the same, a unique solution exists. Physics provides only half of it, so we have to come up with the other half in a way that ensures the vanishing of all \(b_k\) coefficients in Eq.~(\ref{eq:general-solution-for-potential}). The general form of the measurable is
\begin{eqnarray}
u(t) = \sum\limits_{k=1}^Na_kv_k(t)+ w(t),
\label{eq:general-solution-for-observable}
\end{eqnarray}
where \(v_k(t)=\mathcal{\tilde D}\varphi_k(t)\), which is a basis in \(\mathrm{Ker}(\mathcal{D})\).

It is possible to create the initial conditions for the measurable from the initial conditions for the potential.
Let the initial conditions for the potential be
\begin{eqnarray}
\phi_{0,n} = \dv[n-1]{\phi}{t}\bigg|_{t=0} \qq{where} n = \qty{1,\,2,\,\dots 2N},
\end{eqnarray}
and let the initial conditions for the measurable be
\begin{eqnarray}
u_{0,n} = \dv[n-1]{u}{t}\bigg|_{t=0} \qq{where} n = \qty{1,\,2,\,\dots N}.
\end{eqnarray}
The initial conditions for the measurable can be obtained by a linear combination of the initial conditions for the potential. It can be proven by straightforward calculation:
\begin{align}
u_{0,n} &=\dv[n-1]{t} \mathcal{\tilde D}\phi\bigg|_{t=0}=\left[\dv[n-1]{t} \sum\limits_{i=0}^N (-1)^i\dv[i]{t} (p_i\phi)\right]_{t=0} = \nonumber\\[6pt]
&=\left[\dv[n-1]{t} \sum\limits_{i=0}^N\sum\limits_{l=0}^i (-1)^i\binom{i}{l}\dv[l]{t}p_i\cdot \dv[i-l]{t}\phi\right]_{t=0}= \nonumber\\[6pt]
%
%
&=\sum\limits_{i=0}^N\sum\limits_{l=0}^i\sum\limits_{m=0}^{n-1} T_{n,i,l,m}\phi_{0,n+i-l-m},
\label{eq:pot-ic-meas-ic-conn}
\end{align}
where
\begin{eqnarray}
T_{n,i,l,m}=(-1)^i\binom{i}{l}\binom{n-1}{m}\dv[l+m]{t}p_i\bigg|_{t=0}.
\end{eqnarray}
Unsurprisingly, this relation cannot be inverted, but it provides a limitation on the configuration of the potential initial conditions. 

One possible (but not effective) way to find correct initial conditions for the potential in a numerical simulation is to try random configurations which reproduce the physical initial conditions (this can be checked using Eq.~(\ref{eq:pot-ic-meas-ic-conn})). The closer the system starts in the phase space to the configuration that ensures the vanishing of the non-physical part, the slower the divergent part of the solution will start to dominate. Other than trying, it seems improbable, that there is a method to create the desired initial conditions.

\subsection{Theoretical background of higher order Lagrangian and Hamiltonian mechanics}
Usually we are dealing with systems that can be described by a Lagrangian function of the form $L(t, q_i, \dot q_i)$, so it contains at maximum first order derivatives. If we wish to use the method of abstract potential, we need to generalize the Lagrangian formalism to Lagrangians depending on higher order derivatives. We may examine such Lagrangian $L(t, q_i, \dot q_i, \ddot q_i, \dots)$, in which case we can generalize the variational principle \cite{gelfandfomincalculus} 
(Hamilton principle demands that the action functional $S$ is extremal on physical trajectories)
\begin{equation}
0 = \fdv{S}{q} = \fdv{q} \int\limits_{t_1}^{t_2}L(t, q_i, \dot q_i, \ddot q_i, \dots) \dd t
\end{equation}
to obtain equations of motion
\begin{equation}
0 = \sum\limits_{n=0}^N(-1)^n\dv[n]{t}\pdv{L}{\left(\dv[n]{t} q_i\right)}.
\end{equation}
We can also build a Hamiltonian formalism by correctly choosing canonical coordinate and momentum pairs
\begin{subequations}
\begin{align}
q_{i,n}&:=\dv[n-1]{t}q_i\\
p_{i,n}&:=\sum\limits_{k=0}^{N-n}(-1)^k\dv[k]{t}\pdv{L}{\qty(\dv{t}q_{i,n+k})},
\end{align}
\end{subequations}
where \(n = 1,\,\dots\,N\). In that case the Hamiltonian function can be calculated from the Lagrangian function
\begin{equation}
H = \sum\limits_{i}\left(p_{i,1}\dv{q_{i,1}}{t} + p_{i,2}\dv{q_{i,2}}{t} + \dots + p_{i,N}\dv{q_{i,N}}{t}\right)-L
\label{eq:calculate-hamiltonian}
\end{equation}
and we can obtain the canonical equations, which take the usual form
\begin{subequations}
\begin{align}
\dv{q_{i,n}}{t} &= \pdv{H}{p_{i,n}}\\[6pt]
\dv{p_{i,n}}{t} &= - \pdv{H}{q_{i,n}}.
\end{align}
\end{subequations}
As we can see, in the Hamiltonian formalism there are no higher order derivatives, the canonical equations are first order and the dimension of the phase space $M\cdot N$, where $M$ is the number of general coordinates and $N$ is the order of the highest order derivative present in the Lagrangian.

\subsubsection*{Field theoretical generalization}
We could go one step further by generalizing this to Lagrangian densities hence be able to describe field theories of higher order. In order to do this we first define the $N$th order Lagrangian density:
\begin{equation}
L = \int\limits_V\dd x_1\dots\dd x_d\,\mathcal{L}\qty(\phi_i,\, \pdv{\phi_i}{x_k},\, \dots \frac{\partial^N \phi_i}{\partial x_{k_1}\dots x_{k_N}}).
\end{equation}
The $x_0$ coordinate is used as time. The Euler--Lagrange equation is again calculated from the Hamilton principle using variational calculus \cite{courant1954methods}:
\begin{equation}
0 = \sum\limits_{n=0}^{N}\sum\limits_{\alpha_1,\dots\alpha_n = 0}^{d}(-1)^n\frac{\partial^n}{\partial x_{\alpha_1}\dots \partial x_{\alpha_n}}\pdv{\mathcal{L}}{\phi^{(n)}_{i[\alpha_1,\alpha_2,\dots \alpha_n]}},
\end{equation}
where
\begin{equation}
\phi^{(n)}_{i[\alpha_1,\alpha_2,\dots \alpha_n]}=\frac{\partial^n\phi_i}{\partial x_{\alpha_1}\dots \partial x_{\alpha_n}}.
\end{equation}
Now, we introduce the canonical momentum density and canonical field pairs:
\begin{subequations}
\begin{align}
\phi_{i,n}&:=\pdv[n-1]{t}\phi_i\\
\pi_{i,n}&:=\sum\limits_{k=0}^{N-n}(-1)^k\pdv[k]{t}\pdv{\mathcal L}{(\pdv{t}\phi_{i,n+k})}.
\end{align}
\end{subequations}
The Hamiltonian density can be obtained similarly to Eq.~(\ref{eq:calculate-hamiltonian}):
\begin{equation}
\mathcal H = \sum\limits_{i}\left(\pi_{i,1}\pdv{\phi_{i,1}}{t} + \dots + \pi_{i,N}\pdv{\phi_{i,N}}{t}\right)-\mathcal L
\label{eq:calculate-hamiltonian-density}
\end{equation}

\section{The damped linear harmonic oscillator (a toy model)}
The damped harmonic oscillator is a really good toy model, to test different methods on it. The undamped harmonic oscillator is a well-known system both classically, and quantum mechanically, so it provides a good starting point for introducing the damping. The equation of motion for the damped harmonic oscillator is
\begin{equation}
m\ddot x + 2m\lambda \dot x + m\omega^2 x = 0,
\end{equation}
where \(m\) is the mass, \(\lambda\) is the damping coefficient and \(\omega\) is the angular frequency.

In order to define a potential \(q\) for the measurable quantity \(x\), the adjoint equation must be calculated first. As the coefficients are constant, this can be easily done, and the definition equation can be obtained
\begin{equation}
x = \ddot q - 2\lambda \dot q + \omega^2 q.
\label{eq:dlho:potencial_definition}
\end{equation}
By following the method, described in section~\ref{sec:abstract-potential}, the following Lagrangian is received:
\begin{equation}
L = \frac{1}{2} \left( \ddot q - 2\lambda \dot q + \omega^2 q \right)^2.
\end{equation}
The method guarantees, that the Euler--Lagrange equation will be
\begin{equation}
\qty(\dv[2]{t}+2\lambda \dv{t} + \omega^2)\qty(\dv[2]{t}-2\lambda \dv{t} + \omega^2)q = 0.
\label{eq:dlho:lagrange_EOM}
\end{equation}

\subsection{Underdamped and overdamped cases}
As the coefficients are constants in the differential operator, it will commute with its adjoint, which means, that the solution for \(q(t)\) can be easily calculated
\begin{equation}
q(t) = a_1\mathrm{e}^{-(\lambda+\gamma)t} + 
a_2\mathrm{e}^{-(\lambda-\gamma)t}+
b_1\mathrm{e}^{(\lambda+\gamma)t}+
b_2\mathrm{e}^{(\lambda-\gamma)t},
\label{eq:dlho:potential_solution}
\end{equation}
where \(\gamma = \sqrt{\lambda^2-\omega^2}\). The terms proportional to \(\mathrm{e}^{\lambda t}\) are solutions of the adjoint operator, hence they are non-physical solutions, that will not contribute to the measurable \(x(t)\). The effect of the adjoint operator on the other two terms is just a multiplication by a constant value, so they are the two independent solutions of the original differential operator. 

Physics provides the initial conditions for the measurable
\begin{align}
x(t=0) = x_0,\\[4pt]
\dot x(t=0) = v_0.
\end{align}
By choosing the initial conditions for the potential
\begin{align}
q(0) &= \frac{2\lambda x_0+v_0}{4\lambda(\lambda^2-\gamma^2)},\\[4pt]
\dot q(0) &= -\frac{x_0}{4\lambda},\\[4pt]
\ddot q(0) &= -\frac{v_0}{4\lambda},\\[4pt]
\dddot q(0) &= \frac{(\lambda^2-\gamma^2)x_0+2\lambda v_0}{4\lambda},
\end{align}
the non-physical solutions (the exponentially increasing terms in Eq.~(\ref{eq:dlho:potential_solution})) will vanish, so the coefficients will be
\begin{align}
a_1 &= \frac{(\gamma-\lambda)x_0-v_0}{8\gamma\lambda(\lambda+\gamma)},\\[4pt]
a_2 &= \frac{(\gamma+\lambda)x_0+v_0}{8\gamma\lambda(\lambda-\gamma)},\\[4pt]
b_1 &= 0,\\[4pt]
b_2 &= 0.
\end{align}

\subsection{Critical damping and undamped case}
There are 2 interesting cases, when the characteristic equation of the differential equation Eq.~(\ref{eq:dlho:lagrange_EOM}) has repeated roots, \(\lambda=\omega\) and \(\lambda=0\). For \(\lambda=\omega\) the equation of motion is
\begin{equation}
\dv[4]{q}{t}-2\omega^2\dv[2]{q}{t} + \omega^4 q=0,
\end{equation}
for which the general solution and the measurable are
\begin{equation}
q=c_1 \mathrm{e}^{-\omega t}+
c_2 t \mathrm{e}^{-\omega t}+
c_3 \mathrm{e}^{\omega t}+
c_4 t \mathrm{e}^{\omega t},
\end{equation}
\begin{equation}
x=\mathrm{e}^{-\omega t} (4 c_1\omega^2 - 4c_2 \omega + 4 c_2 \omega^2 t).
\end{equation}
Here, the terms proportional to \( \mathrm{e}^{\omega t} \) will not contribute to the measurable, so they will not carry any physical information. This is similar to the previous cases where the exponentially increasing terms were solutions of the adjoint operator. This means, that only the decreasing terms are enough to construct a potential carrying all physical information. We can choose the initial conditions for the potential the following way
\begin{align}
q(0) &= \frac{2\omega x_0+v_0}{4\omega^3},\\[4pt]
\dot q(0) &= -\frac{x_0}{4\omega},\\[4pt]
\ddot q(0) &= -\frac{v_0}{4\omega},\\[4pt]
\dddot q(0) &= \frac{\omega x_0+2 v_0}{4},
\end{align}
it will ensure the vanishing of the non-physical solutions, and will result in the following values of the coefficients \(c_i\)
\begin{align}
c_1 &= \frac{2\omega x_0+v_0}{4\omega^3},\\[4pt]
c_2 &= \frac{\omega x_0 + v_0}{4\omega^2},\\[4pt]
c_3 &= 0,\\[4pt]
c_4 &= 0.
\end{align}

Interestingly, something unexpected occurs, if the \(\lambda=0\) case is investigated. The equation of motion for this special case is
\begin{equation}
\dv[4]{q}{t}+2\omega^2\dv[2]{q}{t} + \omega^4 q=0,
\end{equation}
for which the general solution and the measurable are
\begin{equation}
q=c_1 \mathrm{e}^{-i\omega t}+
c_2 t \mathrm{e}^{-i\omega t}+
c_3 \mathrm{e}^{i\omega t}+
c_4 t \mathrm{e}^{i\omega t},
\end{equation}
\begin{equation}
x=-c_2 2i\omega \mathrm{e}^{-i\omega t} + c_4 2i\omega \mathrm{e}^{i\omega t}.
\end{equation}
As it can be seen, only the polynomially increasing terms carry physical information. This might lead to the assumption, that if information is encoded in increasing terms of the general solution, the system is not dissipative. However, the validity of this assumption is a question.

In this case, it is also possible to choose the initial conditions, so the non-physical terms will vanish. The correct choice is
\begin{align}
q(0) &= 0,\\[4pt]
\dot q(0) &= -\frac{v_0}{2\omega^2},\\[4pt]
\ddot q(0) &= x_0,\\[4pt]
\dddot q(0) &= \frac{3v_0}{2},
\end{align}
with which the coefficients \(c_i\) are
\begin{align}
c_1 &= 0,\\[4pt]
c_2 &= \frac{-v_0 + i \omega x_0}{4\omega^2},\\[4pt]
c_3 &= 0,\\[4pt]
c_4 &= \frac{-v_0 - i \omega x_0}{4\omega^2}.
\end{align}

\section{Conclusion}
By creating a potential, linear differential equations describing a dissipative system can be calculated from a Lagrangian. Using the described method, the potential can be easily constructed to an equation that expresses the dissipative behaviour of a a physical quantity. The problem of properly modelling the environment vanishes, instead the adjoint equation (which defines the connection between the potential and the measurable) must be solved. Although in concept, it is possible to omit non-physical solution which result in instabilities, technically there is no way to do that if the equation cannot be solved analytically. Moreover, if initial conditions, that provide a zero non-physical part in the solution, are found, during a numerical simulation, numerical errors can result in an unstable solution. One proper way to stabilize such a simulation is to find a relation that can be checked throughout the solving procedure and restricts the solution to the physical part only. 

The benefit of this method not only lies in the fact, that it provides a way of receiving an equation from a Lagrangian, but it provides the powerful tools of the Lagrangian framework. Of course, it is an open question how handy these tools are on the level of potentials, and how the physical information is obtained. One highly interesting idea is quantization, whether it is even possible through this method. Another exciting utilization is coupling fields\citep{MARKUS20092122}.


%
%

%

\begin{acknowledgments}
Support by the Hungarian National Research, Development and Innovation
Office of Hungary (NKFIH) Grant Nr. K119442 is acknowledged.
\end{acknowledgments}

\section*{Data availability statement}
Data sharing is not applicable to this article as no new data were created or analyzed in this study.

\bibliographystyle{abbrv}
\bibliography{references}

\end{document}